\documentclass{article}
\usepackage{epsfig}
\usepackage{a4wide}
\def\S12{\mathrm{S}_{1,2}}

\newcommand{\Hp}{$H^+$}
\newcommand{\Gp}{$G^+$}

\newcounter{allequation}
\newcommand{\subnumbers}{\setcounter{allequation}{\value{equation}}
      \addtocounter{allequation}{1}
      \setcounter{equation}{0}
       \renewcommand{\theequation}{\arabic{allequation}\alph{equation}}}
\newcommand{\normalnumbers}{\setcounter{equation}{\value{allequation}}
                       \renewcommand{\theequation}{\arabic{equation}}}

\begin{document}

\title{
  {Mass Terms in Two-Higgs Doublet Models}\\[0.75cm]} 
\author{R. Santos${}^{a,b,1}$, S. M. Oliveira${}^{a,2}$,
A. Barroso${}^{b,3}$\\[1.25cm]
  {\em ${}^a$ Centro de F\'\i sica Te\'orica e Computacional,
   Faculdade de Ci\^encias, Universidade de Lisboa,}\\
  {\em  Av. Prof. Gama Pinto 2, 1649-003 Lisboa, Portugal}\\[0.5cm]
  {\em  ${}^b$ Instituto Superior de Transportes e
   Comunica\c{c}\~oes, Campus Universit\'ario}\\
  {\em  R. D. Afonso Henriques, 2330-519 Entroncamento, Portugal}\\[2.0cm]}
\date{\today}
\maketitle
\footnotetext[1]{e-mail: rsantos@cii.fc.ul.pt}
\footnotetext[2]{e-mail: smo@cii.fc.ul.pt}
\footnotetext[3]{e-mail: barroso@cii.fc.ul.pt}

\thispagestyle{empty}

\begin{abstract}{
 We take a closer look at the mass terms of all renormalizable and
CP conserving two-Higgs doublet models (THDM). We show how
some of the dimension two parameters in the potential can be
set equal to zero leading to relations among the tree-level
parameters of the potential. The different versions
of the THDM obtained give rise to different amplitudes for
physical processes. We will illustrate this with two examples.
The first one is the one-loop weak correction to the top decay
width, $t \rightarrow b W$. The second one is the decay
$h \rightarrow \gamma \gamma$ in the fermiophobic limit.
}
\end{abstract}
\newpage

\section{Introduction}
The generation of particle masses in the standard model (SM) is
accomplished through the well known Higgs mechanism.
In its minimal version, i.e., with one scalar
doublet, there is only one free parameter in the scalar sector,
the Higgs mass, $M_H$. After spontaneous symmetry breaking
the two parameters in the scalar potential are replaced
by the vacuum expectation value $v=247 \;GeV$
and by $M_H$, which remains unconstrained.
Introducing another scalar doublet of complex 
fields makes the number of free parameters in the potential  grow
from two to fourteen. The number of scalar particles grows
from one to four. If explicit CP violation is not allowed, the number
of free parameters is reduced to ten. This number
can be further reduced to seven by imposing either a global U(1)
symmetry or a $Z_2$ symmetry. We call the resulting
potentials, $V_A$ and $V_B$ respectively. Some of the vertices
in the scalar sectors of $V_A$ and $V_B$ are different
and this gives rise to differences
in the amplitudes for physical processes.
This fact has often been overlooked in the literature. 

Adding or subtracting mass terms to either of these
potentials does not spoil the renormalizability
of the model. In the first case it is well known that a softly
broken theory remains
renormalizable (see for instance \cite{IZ}). In the second case
the renormalization will introduce a counterterm with the same
structure of the suppressed term. Again these are mass terms
which are simply absorbed in the mass renormalization of the
four scalar bosons.  
Hence, we have analysed
all possible CP-conserving and renormalizable models obtained
by addition or subtraction of all available mass terms to
$V_A$ and $V_B$. In some cases the resulting potentials have 
only six or even five free parameters. 
In paragraph II we discuss the potential. Then, to illustrate
that they lead to different physical results we calculate
in paragraph III the one-loop correction to the
top quark decay $t \rightarrow b W$
and in paragraph IV the decay $h \rightarrow \gamma \gamma$
in the fermiophobic limit.

\section{The potentials}

To define our notation, we start with a brief review of the
two-Higgs doublet models (THDM). Let $\Phi_i$ with $i=1,2$
denote two complex scalar doublets with hyper-charge 1.  
Under $C$ the fields transform
as $\Phi_i\rightarrow e^{i\alpha_i} \Phi_i^*$ where the parameters
$\alpha_i$ are arbitrary.
Defining the complete set of invariants $x_1 = \phi_1^\dagger \phi_1$,
$x_2 = \phi_2^\dagger \phi_2$, $x_3 = \Re\{\phi_1^\dagger 
\phi_2\}$ and $x_4 = \Im\{\phi_1^\dagger \phi_2\}$
and choosing $\alpha_1=\alpha_2=0$ we can
write the most general $SU(2) \otimes U(1)$ invariant, $C$
invariant and renormalizable potential in the form
\begin{equation}
V=-\mu_{1}^{2} x_{1}-\mu_{2}^{2} x_{2}-\mu_{3}^{2} x_{3}
+\lambda_{1} x_{1}^2+\lambda_{2} x_{2}^2
+\lambda_{3} x_{3}^2+\lambda_{4} x_{4}^2+\lambda_{5} x_{1}
 x_{2} +\lambda_6 x_1 x_3 +\lambda_7 x_2 x_3 \enskip . 
\end{equation}
In a previous paper\cite{VSB94} we have studied the different types
of extrema for potential $V$. We have shown in
\cite{VSB94} that there are two natural ways of imposing that a
minimum with $CP$ violation never occurs, i.e., that the minimum
of the potential is of the form 
\begin{equation}
 \left<\Phi_1\right>  =  \left(\begin{array}{l} 0 \\ v_1
 \end{array}\right)  \qquad \qquad \qquad
 \left<\Phi_2\right>  =  \left(\begin{array}{l} 0 \\ v_2
 \end{array}\right) \quad ,
\end{equation}
with $v_i$ real. The first one, denoted $V_A$, is
obtained by setting $\mu_3^2=\lambda_6=\lambda_7=0$ in
equation (1). This is the potential in Ref. \cite{sher1}.
It is invariant under the $Z_2$ transformation 
$\Phi_1 \rightarrow \Phi_1$ and $\Phi_2 \rightarrow -\Phi_2$.
The second 7-parameter potential, denoted by $V_B$,
is the potential obtained in the Minimal Supersymmetric Model (MSSM)
and it corresponds to the conditions $\lambda_6=\lambda_7=0$
and $\lambda_3=\lambda_4$. Potential $V_B$ is invariant
under a global $U(1)$ symmetry  $\Phi_2 \rightarrow e^{i \alpha}\Phi_2$
except for the soft breaking term proportional to $\mu_{3}^2$.

Let us examine how
the theory behaves by adding or subtracting all possible mass terms
to the potential. We start by writing the potential in the form
\begin{equation}
V_{AB}=-\mu_{1}^{2} x_{1}-\mu_{2}^{2} x_{2}-\mu_{3}^{2} x_{3}
-\mu_{4}^{2} x_{4}+\lambda_{1} x_{1}^2+\lambda_{2} x_{2}^2
+\lambda_{3} x_{3}^2+\lambda_{4} x_{4}^2+\lambda_{5} x_{1}
 x_{2} \enskip , 
\end{equation}
which allow us to study both potentials $V_A$ and $V_B$
at the same time.
The minimum conditions are
\subnumbers
\begin{eqnarray}
 & T_1 = & v_1 \left[-\mu_1^2 + \lambda_1 v_1^2 +\lambda_{+} v_2^2
     \right] - \frac{1}{2} \mu_3^2 v_2 = 0 \\ 
 & T_2 = & v_2 \left[-\mu_2^2 + \lambda_2 v_2^2 +\lambda_{+} v_1^2
     \right] - \frac{1}{2} \mu_3^2 v_1 = 0 \\
 & T_3 = &  \mu_4^2 v_2 = 0 \\
 & T_4 = &  - \mu_4^2 v_1 = 0 \quad ,
\end{eqnarray}
\normalnumbers
with $\lambda_{+} = (\lambda_3 + \lambda_5 )/2$.
Obviously, conditions (4c,d) force $\mu_4^2 = 0$ because
the vacuum configuration chosen in Eq. (2) implies a stable,
CP-conserving minima. Since the $\mu_4^2$ term allows
mixing between all neutral fields, its existence would
only be possible in a theory with spontaneous CP violation.
Each complex doublet \(\phi_i\) can be written as
\begin{equation}
\phi_{i}=
\left[
 \begin{array}{c}
a_{i}^+  \\ (v_{i}+b_{i}+ic_{i})/ \sqrt{2}
 \end{array} \right]
\end{equation}
where \(a_{i}^+\) are complex fields, and \(b_i\) and
\(c_i\) are real fields. This, in turn, enables us to
write the mass terms of potential $V_{AB}$ as:
\begin{eqnarray}
V_{AB}^{mass} & = & 
\left[
\begin{array}{c}
a_{1}^+ \quad a_{2}^+ 
\end{array} \right]
M_{a}
\left[
\begin{array}{c}
a_{1}^- \\ a_{2}^- 
\end{array} \right]
+\frac{1}{2}
\left[
\begin{array}{c}
c_{1} \quad c_{2}
\end{array} \right]
M_{c}
\left[
\begin{array}{c}
c_{1} \\ c_{2} 
\end{array} \right] \nonumber \\ & & \nonumber \\
& & +\frac{1}{2}
\left[
\begin{array}{c}
b_{1}\quad b_{2}
\end{array} \right]
M_{b}
\left[
\begin{array}{c}
b_{1} \\ b_{2} 
\end{array} \right] 
\nonumber \enskip ,
\end{eqnarray}
with the matrices $M_{a}$, $M_{b}$ and $M_c$ defined as
\renewcommand{\theequation}{\arabic{equation}\mbox{a}}
\begin{equation}
M_{a}=\frac{1}{2}
\left[
\begin{array}{cc}
\frac{v_2}{v_1} \mu_3^2 -v_2^2 \lambda_3
&  - \mu_3^2 + v_1 v_2 \lambda_3  \\
 - \mu_3^2 + v_1 v_2 \lambda_3 
& \frac{v_1}{v_2} \mu_3^2 -v_1^2 \lambda_3
\end{array} \right]
\end{equation}
\addtocounter{equation}{-1}
\renewcommand{\theequation}{\arabic{equation}\mbox{b}}
\begin{equation}
M_{b}=
\left[
\begin{array}{cc}
2v_1^2 \lambda_1+ \frac{v_2}{2 v_1} \mu_3^2 
&  v_1 v_2 (\lambda_3  + \lambda_5)- \frac{\mu_3^2}{2} \\
v_1 v_2 (\lambda_3 + \lambda_5)- \frac{\mu_3^2}{2} 
& 2 v_2^2 \lambda_2+ \frac{v_1}{2 v_2} \mu_3^2  
\end{array} \right]
\end{equation}
\addtocounter{equation}{-1}
\renewcommand{\theequation}{\arabic{equation}\mbox{c}}
\begin{equation}
M_{c}= \frac{1}{2}
\left[
\begin{array}{cc}
v_2^2 (\lambda_4 - \lambda_3) + \frac{v_2}{v_1} \mu_3^2 
&  -v_1 v_2 (\lambda_4 - \lambda_3) -  \mu_3^2 \\
- v_1 v_2 (\lambda_4 - \lambda_3) -  \mu_3^2 
&  v_1^2 (\lambda_4 - \lambda_3) + \frac{v_1}{v_2} \mu_3^2 
\end{array} \right] \enskip .
\end{equation}
\renewcommand{\theequation}{\arabic{equation}}
Diagonalizing the quadratic terms of $V_{AB}$ one
obtains the mass eigenstates: 2 neutral CP-even scalar
particles, H and h, a neutral CP-odd scalar particle,
A, the would-be Goldstone boson partner of the Z,
$G_0$, a charged Higgs field, \Hp and the Goldstone
associated with the W boson, \Gp. The relations between
the mass eigenstates and the SU(2)$\otimes$U(1) eigenstates are:
\renewcommand{\theequation}{\arabic{equation}\mbox{a}}
\begin{equation}
\left[
\begin{array}{c}
H \\ h
\end{array} \right]
=R_{\alpha}
\left[
\begin{array}{c}
b_1 \\ b_2
\end{array} \right] \qquad
\left[
\begin{array}{c}
H^+ \\ G^+
\end{array} \right]
=R_{\beta}
\left[
\begin{array}{c}
a_1^+ \\ a_2^+
\end{array} \right] \qquad
\left[
\begin{array}{c}
A \\ G_0
\end{array} \right]
=R_{\beta}
\left[
\begin{array}{c}
c_1 \\ c_2
\end{array} \right]
\end{equation}
with
\addtocounter{equation}{-1}
\renewcommand{\theequation}{\arabic{equation}\mbox{b}}
\begin{equation}
R_{\alpha}=
\left[
\begin{array}{cc}
\cos \alpha & \sin \alpha \\
- \sin \alpha & \cos \alpha
\end{array} \right] \qquad
R_{\beta}=
\left[
\begin{array}{cc}
- \sin \beta & \cos \beta \\
\cos \beta & \sin \beta
\end{array} \right] \enskip ,
\end{equation}
and $0 < \beta < \frac{\pi}{2}$,
$-\frac{\pi}{2} < \alpha < \frac{\pi}{2}$. 
The particle masses and the angles are given by
\subnumbers
\begin{eqnarray}
& & M_{H^+}^2 =  -\frac{1}{2} v^2 \left[ \lambda_3 - \frac{\mu_{3}^2}
              {v_1 v_2} \right] \\
& & M_{A}^2 =  \frac{1}{2} \mu_{3}^2
      \,\frac{v^2}{v_1 v_2} +\frac{1}{2} (\lambda_4-\lambda_3) v^2 \\ 
& & M_{H,h}^2 =  \lambda_1 v_1^2 + \lambda_2 v_2^2 + \mbox{$\frac{1}{4}$} 
     \mu_{3}^2\left(\mbox{$\frac{v_2}{v_1}$}+\mbox{$\frac{v_1}{v_2}$}\right)\,
    \\ & & \,  
   \qquad \qquad \pm \sqrt{ \left( \lambda_1 v_1^2 - \lambda_2 v_2^2 + \mbox{$\frac{1}{4}$} 
     \mu_{3}^2\left(\mbox{$\frac{v_2}{v_1}$}-\mbox{$\frac{v_1}{v_2}$}\right) 
     \right)^2 + \left( v_1 v_2 (\lambda_3 + \lambda_5) - \frac{1}{2}
                 \mu_{3}^2\right)^2 }  \\ 
& & \tan 2\alpha = \frac{2\,v_1
  v_2 \,\lambda_+ -\frac{1}{2} \mu_{3}^2}{\lambda_1 v_1^2 - \lambda_2 v_2^2
  + \frac{1}{4} \mu_{3}^2 
  \left(\mbox{$\frac{v_2}{v_1}$}-\mbox{$\frac{v_1}{v_2}$}\right)} \\
& & \tan \beta = \frac{v_2}{v_1} \enskip .
\end{eqnarray}
\normalnumbers
To derive the cubic and quartic vertices stemming from the potential
it is convenient to rewrite the $\lambda_i$ in terms of the masses
and angles.The results are:
\subnumbers
\begin{eqnarray}
\lambda_1 & = & \frac{1}{2v^2 \cos^2 \beta} \left[ M_H^2 \cos^2 \alpha +
    M_h^2 \sin^2 \alpha - \frac{\mu_{3}^2}{2} \tan \beta \right] \\
\lambda_2 & = &  \frac{1}{2v^2 \sin^2 \beta} \left[ M_H^2 \sin^2 \alpha +
    M_h^2 \cos^2 \alpha - \frac{\mu_{3}^2}{2} \tan^{-1} \beta \right] \\ 
\lambda_3 & = & - \frac{2}{v^2} \left[ M_{H^+}^2 -
     \frac{\mu_{3}^2}{\sin (2 \beta)} \right] \\ 
\lambda_4 & = & \frac{2}{v^2} \left[ M_A^2 - M_{H^+}^2 \right] \\
\lambda_5 & = & \frac{\sin (2 \alpha)}{v^2 \sin (2 \beta)} \left[ M_H^2 -
     M_h^2 \right] + \frac{2}{v^2} M_{H^+}^2 - 
     \frac{\mu_{3}^2}{v^2 \sin (2 \beta)} \enskip . 
\end{eqnarray}
\normalnumbers
\subsection{Potential $V_A$}
Setting $\mu_4^2 = 0$ we obtain potential $V_A$
with the extra soft breaking term
$- \mu_3^2 x_3$, i.e.,
\begin{equation}
V_{A}=-\mu_{1}^{2} x_{1}-\mu_{2}^{2} x_{2}-\mu_{3}^{2} x_{3}
+\lambda_{1} x_{1}^2+\lambda_{2} x_{2}^2
+\lambda_{3} x_{3}^2+\lambda_{4} x_{4}^2+\lambda_{5} x_{1}
 x_{2} \enskip . 
\end{equation}
We will now consider the different possibilities regarding the
mass terms of the potential. Obviously, the conditions 
$\mu_{1}^{2} = \mu_{2}^{2} = \mu_{3}^{2} = 0$ would lead
to a potential with no mass scale. This particular case was
studied by Coleman and Weinberg \cite{CW73} for the SM
and by Inoue \textit{et al} for the THDM \cite{IKN80}. In this model
the scalars acquire a mass as a result of radiative corrections.
Hence, we end up with a renormalizable model where all scalars
have a definite mass which is related to the gauge bosons masses.
If $\mu_{3}^{2} = 0$ one obtains a seven parameter potential
where all vertices are given in terms of the physical masses and
the angles $\alpha$ and $\beta$. However, if $\mu_{3}^{2} \neq 0$, 
Eqs. (9) show that there are additional contributions to the
cubic and quartic scalar vertices proportional
to this parameter. In fact, now, the
potential has eight free parameters.
The inclusion of this extra
parameter can change the tree-level unitarity bounds obtained
for $V_A$ \cite{AAN00}.

When just one mass term is set equal to
zero, we obtain the potential $V_A$ described in
\cite{SB97}. Hence, the most 
interesting cases are the ones where just one mass
terms is different from zero.
Potentials with two zero mass terms have
the same number of particles, the same number of rotation
angles and the same number of vertices than the original
potential. However, at tree-level, the seven parameters in the potential
are no longer free. In fact, the number of free parameters
is reduced to six by one of the following relations:
\begin{enumerate}
\item[i)] $\mu_1^2 = \mu_3^2 = 0$

\renewcommand{\theequation}{\arabic{equation}\mbox{a}}
\begin{equation}
\tan \beta = - 2 \frac{M_H^2 \cos^2 \alpha + M_h^2 \sin^2 \alpha}
{\sin (2 \alpha) (M_H^2 - M_h^2)} \enskip , 
\end{equation}
\addtocounter{equation}{-1}

\item[ii)] $\mu_2^2 = \mu_3^2 = 0$

\renewcommand{\theequation}{\arabic{equation}\mbox{b}}
\begin{equation}
\tan \beta = - \frac{1}{2} \sin (2 \alpha) 
\frac{M_H^2 - M_h^2}
{M_H^2 \sin^2 \alpha + M_h^2 \cos^2 \alpha} \enskip , 
\end{equation}
\addtocounter{equation}{-1}

\item[iii)] $\mu_1^2 = \mu_2^2 = 0$

\renewcommand{\theequation}{\arabic{equation}\mbox{c}}
\begin{equation}
\tan^2 \beta = \frac{M_H^2 \cos^2 \alpha + M_h^2 \sin^2 \alpha}
{M_H^2 \sin^2 \alpha + M_h^2 \cos^2 \alpha} \enskip . 
\end{equation}
\renewcommand{\theequation}{\arabic{equation}}

\end{enumerate}
As we have already pointed out, these relations are not valid
beyond tree-level. In fact, the renormalization scheme forces
us to treat the angles and the masses as independent parameters.

We denote these models by $V_A^{i}$, $V_A^{ii}$ and $V_A^{iii}$.
In $V_A^{i}$, $V_A^{ii}$
one can readily see that $\sin(2 \alpha) < 0$, that is, the
rotation angle $\alpha$ is now bounded to be in the interval
$-\frac{\pi}{4}< \alpha <0$. This, in turn, implies that
all Yukawa couplings have a definite sign. In $V_A^{iii}$
$\tan \beta$ is in the range
\begin{equation}\label{tanbeta}
\frac{M_h}{M_H} < \tan \beta < \frac{M_H}{M_h} \enskip , 
\end{equation}
which means that values of $\tan \beta$ close to zero or infinity
are not allowed.

In the limit $M_H >> M_h$ all equations (11,a,b,c)
are reduced to
\begin{equation}\label{tbetatalfa}
\tan \beta \approx - \frac{1}{\tan \alpha} \enskip . 
\end{equation}
This condition implies that all couplings between the fermions
and the lightest Higgs, $h$, are the SM ones.
The remaining Yukawa couplings will be proportional
to either $\tan \beta$ or $\cot \beta$.
The couplings between the scalars and the gauge bosons
are such that $\cos (\alpha - \beta) \approx 0$
and $\sin (\alpha - \beta) \approx -1$.
Hence, in this limit
the lightest Higgs, $h$, resembles very much the 
SM Higgs boson. It can only be distinguished
from the SM Higgs through the couplings to the 
other scalars. On the other hand
the heavier CP-even Higgs, $H$, couples very weekly to
the gauge bosons- it is almost bosophobic.

In the limit where the CP-even Higgs are almost
degenerated ($M_H \approx M_h$) the values of $\tan \beta$
can be quite different in the three models. In $V_A^{i}$ $\tan \beta$
tends to be very large while in $V_A^{ii}$ it is close to zero. 
Finally, in $V_A^{iii}$  $\tan \beta \approx 1$.

\subsection{Potential $V_B$}
Setting $\lambda_3 = \lambda_4$ and $\mu_3^2 \neq 0$ we 
obtain potential $V_B$, i.e.,
\begin{equation}
V_{B}=-\mu_{1}^{2} x_{1}-\mu_{2}^{2} x_{2}-\mu_{3}^{2} x_{3}
+\lambda_{1} x_{1}^2+\lambda_{2} x_{2}^2
+\lambda_{3} \left[ x_{3}^2+ x_{4}^2 \right]+\lambda_{5} x_{1}
 x_{2} \enskip , 
\end{equation}
Unlike $V_A$, when all mass terms are different from zero there
is no extra parameter. Again, if all mass terms are
zero, masses are generated through radiative corrections as
in potential $A$. If $\mu_3^2 = 0$ and $\mu_{1,2}^2 \neq 0$ the 
CP-odd scalar becomes massless at tree-level. 

Forcing $\mu_3^2 \neq 0$ we obtain the following relation:
\begin{enumerate}
\item[i)] $\mu_1^2  = 0$

\renewcommand{\theequation}{\arabic{equation}\mbox{a}}
\begin{equation}
M_A^2 = \frac{1}{2 \sin^2 \beta} \left[ 
M_H^2 \cos^2 \alpha + M_h^2 \sin^2 \alpha + 
\frac{1}{2} \sin (2 \alpha) \tan \beta (M_H^2 - M_h^2) \right] \enskip , 
\end{equation}
\addtocounter{equation}{-1}

\item[ii)] $\mu_2^2 = 0$

\renewcommand{\theequation}{\arabic{equation}\mbox{b}}
\begin{equation}
M_A^2 = \frac{1}{2 \cos^2 \beta} \left[ 
M_H^2 \sin^2 \alpha + M_h^2 \cos^2 \alpha + 
\frac{\sin (2 \alpha)}{2 \tan \beta} (M_H^2 - M_h^2) \right] \enskip ,
\end{equation}
\addtocounter{equation}{-1}

\item[iii)] $\mu_1^2 = \mu_2^2 = 0$

\renewcommand{\theequation}{\arabic{equation}\mbox{c}}
\begin{equation}
\tan^2 \beta = \frac{M_H^2 \cos^2 \alpha + M_h^2 \sin^2 \alpha}
{M_H^2 \sin^2 \alpha + M_h^2 \cos^2 \alpha} \enskip , 
\end{equation}
and either Eq. (15a) or (15b). In this last case, at tree-level,
the number
of free parameters in the potential is reduced from
seven to five.
\renewcommand{\theequation}{\arabic{equation}}
\end{enumerate}
We denote these models by $V_B^{i}$, $V_B^{ii}$ and $V_B^{iii}$.
Similarly to the previous case, Eqs. (15a) and (15b) could also
imply a restriction on the allowed values of $\alpha$. In fact,
since $M_A > 0$, depending on the values of $M_H$, $M_h$ and
$\tan \beta$, some values of  $\alpha$ can be excluded.
For the sake of simplicity, we have chosen $M_A$ as the dependent
variable in these two relations. If a process does not depend
heavily on the value of $M_A$, it is likely that $V_B^{i}$
and $V_B^{ii}$ give the same result as the original $V_B$.

Let us now look at $V_B^{iii}$. Equation (\ref{tanbeta})
obtained for $V_A$ still
holds in this case. In the limit $M_H >> M_h$ we have
\begin{equation}
\tan \beta \approx \frac{1}{\tan \alpha} \quad ;
\enskip M_A \approx M_H \enskip
\end{equation}  
and $0 < \alpha < \pi/2$. All conclusions on the behaviour
of the lightest Higgs boson for $V_A$ also apply in this case.
On the other hand, if $M_H \approx M_h$, then
\begin{equation}
\tan \beta \approx 1 \quad ;
\enskip M_A \approx M_H \approx M_h \enskip
\end{equation}  
and $- \pi/2 < \alpha < \pi/2$.

\section{The Decay $\lowercase{t} \longrightarrow \lowercase{b} \, W^+$}

The electroweak corrections to the decay
$t \rightarrow b W^+$ of order $\alpha^2$ were evaluated
in Refs. \cite{DS91} and \cite{EMMS91}. However, the
renormalization of the Cabibbo-Kobayashi-Maskawa (CKM)
matrix was done in such a way that the final result was gauge
dependent \cite{GGM99}. Recently, using a renormalization prescription
introduced before \cite{BBS00}, we have evaluated this
decay width \cite{OBSB01} in the SM. A similar
calculation in the THDM was done by  Denner and Hoang \cite{DH93},
but again it suffers from the same gauge dependence anomaly.
Despite the fact that the contributions of the  $\delta V_{tb}$
counterterm is small we repeat here the calculation \cite{Sancho}
using the renormalization prescription proposed in Refs. \cite{SB97}
and \cite{BBS00}. Skipping the details of the calculation
that can be found in Ref. \cite{OBSB01}, we present the results
in terms of the parameter
\begin{equation}
\delta \, = \, \frac{2 \, \Re [T_0 T_1^{+}]}{\mid T_0 \mid^2} \enskip , 
\end{equation}
where $T_0$ and $T_1$ are the tree-level and one-loop amplitudes
respectively. Hence, including one-loop corrections, the decay width is
\begin{equation}
\Gamma_1 \, = \, \Gamma_0 \, [1  + \delta ] \enskip ,
\end{equation}
where $\Gamma_0$ is the tree-level decay width.
\\[-0.1cm]
\begin{figure}[htbp]
  \begin{center}
    \epsfig{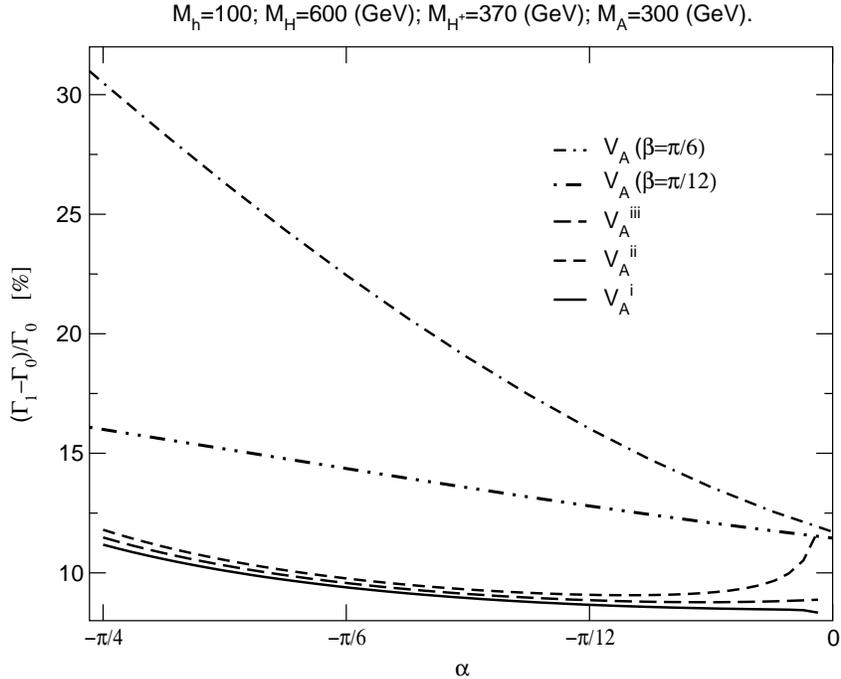}
    \caption{$\delta$ as a function of $\alpha$
    for models $V_A$ and $V_A^{i,ii,iii}$}
    \label{fig_VA}
  \end{center}
\end{figure}
The parametric space of the THDM is very large. Hence, it is not
possible to show the results for a systematic scan of this space.
It is more interesting to present the qualitative trend and to
illustrate the differences that occur for the different versions
of the model. In general $\delta$ is larger than in the SM by a
factor of order two or three. In Fig. 1 we show $\delta$ as 
a function of $\alpha$ for the models of type A, using $M_h=100 \, GeV$
and $M_H=600 \, GeV$. In this case, the three models $V_A^{i,ii,iii}$
give almost the same result. This is a simple consequence that they
are not far from the limit given by Eq. (13). On the same figure
we also plot the value of $\delta$ calculated with $V_A$ for two
values of $\beta$. Now, it is possible to obtain results that are
almost one order of magnitude larger than in the SM
($\delta_{SM} = 4.46 \%$ \cite{OBSB01}). Furthermore, this
is possible for values of masses and angles that do not give
a large contribution to $\rho = M_W^2/(M_Z^2 \cos^2 \theta_W)$
due to the Higgs bosons. We only consider regions of the
parameter space which give $|\Delta \rho| < 2 \times 10^{- 3}$.

In Fig. 2 we plot $\delta$ for type B models. In most cases
values of $\delta$ of the order of $10 \%$ are obtained,
but one does not need to go to an extreme corner of the parametric
space to obtain even larger values. The curves for models
$V_B^{i,ii,iii}$ are not plotted in the entire range of $\alpha$.
This is due to two reasons. Some values of $\alpha$ are excluded
by Eqs. (15) while others are excluded by our  $\Delta \rho$
criteria.

One should notice that for the top decay there are no contributions
due to the cubic scalar vertices. This means that, for this
process, $V_A$ and $V_B$ give exactly the same results. Hence,
the differences illustrated in Figs. 1 and 2 are entirely due
to the restrictions induced by Eqs. (11) and (15).
\\
\begin{figure}[htbp]
  \begin{center}
    \epsfig{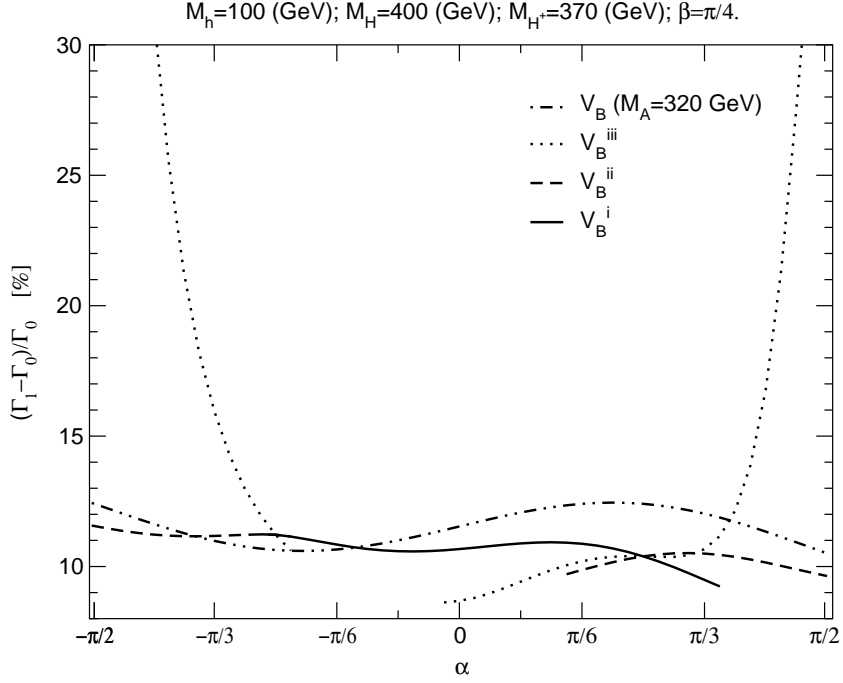}
    \caption{$\delta$ as a function of
     $\alpha$ for models $V_B$ and $V_B^{i,ii,iii}$. In model $V_B^{iii}$
     $\beta$ is given by Eq. (15c)}
    \label{fig_VB}
  \end{center}
\end{figure}

\section{The fermiophobic limit}
In the so-called
fermiophobic limit \cite{DW94,BBS99}, the lightest
Higgs decouples from the fermions at tree level.
This is accomplished in the THDM by setting $\alpha=\pi/2$ in model
I, as defined in Ref. \cite{SB97}.
In Ref. \cite{BS00} we have studied all possible decays of a fermiophobic
Higgs. The dominant decay, for the LEP energies is
$h \rightarrow \gamma \gamma$ \cite{Sofia1}. We have showed
before \cite{BBS99} that
this decay width is different in models $A$ and $B$ due to 
a difference in the $h H^+ H^-$ vertex. Similarly the same
is true for some of the models
derived from $V_A$ and $V_B$. 
Models $V_A^i$ and $V_A^{ii}$ do not allow a fermiophobic Higgs. 
In  $V_A^i$ $\alpha = \pi/2$ leads to $M_h=0$ and in $V_A^{ii}$
it forces $M_H = 0$. In  $V_A^{iii}$ equation (11c) is reduced to
\begin{equation}
\tan \beta = \frac{M_h}{M_H} \enskip .
\end{equation}
Then
\begin{equation}
\frac{1}{2} < \sin^2 (\alpha - \beta) < 1 \enskip
\end{equation} 
where the value $\sin^2 (\alpha - \beta) \approx 1$ is obtained
when  $M_h << M_H$. 

A detailed study of this process using either $V_A$ or $V_B$
was done before \cite{BBS99}. So there is no need to repeat
it here. What remains to be done
is to discuss the changes induced by
the more constrained potentials  $V_A^{i,ii,iii}$ and
$V_B^{i,ii,iii}$.
To do this we denote by $R_A$ the ratio of the
$h \rightarrow \gamma \gamma$ widths calculated in models
$V_A^{iii}$ and  in model $V_A$. For small values 
of $\beta$  $R_A$ is of the order one. However, for
larger $\beta$, $\beta = \pi/3$ for instance,  $R_A \approx 4$.
This could be relevant when analysing the experimental data
to exclude some mass regions.

In potential $B$ a fermiophobic Higgs is always allowed.
In this limit Eqs. (15) can be written as
\begin{enumerate}
\item[i)] $V_B^{i}$
\renewcommand{\theequation}{\arabic{equation}\mbox{a}}
\begin{equation}
\tan^2 \beta = \frac{M_h^2}{2 M_A^2-M_h^2} \enskip , 
\end{equation}
\addtocounter{equation}{-1}

\item[ii)] $V_B^{ii}$
\renewcommand{\theequation}{\arabic{equation}\mbox{b}}
\begin{equation}
\tan^2 \beta = \frac{2 \, M_A^2-M_H^2}{M_H^2} \enskip ,
\end{equation}
\addtocounter{equation}{-1}

\item[iii)]  $V_B^{iii}$
\renewcommand{\theequation}{\arabic{equation}\mbox{c}}
\begin{equation}
\tan \beta = \frac{M_h}{M_H} \quad , \qquad M_A^2 = \frac{M_H^2+M_h^2}{2} 
\end{equation}
\\
\renewcommand{\theequation}{\arabic{equation}}
\end{enumerate}
\begin{figure}[htbp]
  \begin{center}
    \epsfig{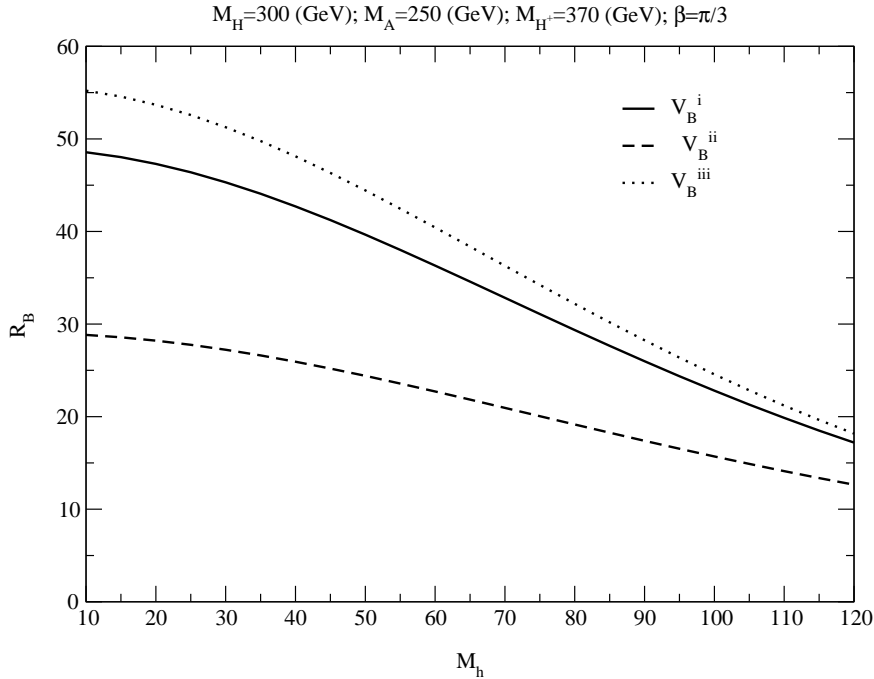}
    \caption{$R_B$ as a function of $M_h$ for $\beta= \pi/3$}
    \label{fig_VB}
  \end{center}
\end{figure}

Again, defining similar ratios $R_B$ for potentials B,
we show in Fig. 3 their variation with $M_h$, for $\beta = \pi/3$.
Now the enhancement is even larger. On the contrary, for
smaller values of $\beta$,  $\beta = \pi/12$ for instance, $R_B$
varies between 0.7 e 1.2. 

\section{Conclusion} 

We have studied the different versions of a renormalizable
and CP-conserving THDM. In general the
different potentials of these models fall into two main
categories. They are either invariant under $Z_2$ or under
a global $U(1)$. In any of these cases the symmetry can
be softly broken with the addition of mass terms.
Alternatively if some of the mass terms are set to zero
one obtains constraining relations among the parameters
of the potential. Despite the fact that these models
are very similar they can give substantial different
predictions for the amplitude of physical processes.
This point is illustrated with two examples. The first
one is a calculation of the one-loop weak corrections
to the top quark decay $t \rightarrow b W$. In general
the THDM give loop contributions that are larger than in the
SM by a factor of two or three. For some values of the
parameters, it is even possible to obtain one order
of magnitude enhancement of the SM result. The second
example is the decay $h \rightarrow \gamma \gamma$ for
a fermiophobic Higgs. Now the results are even more
sensitive to the different potentials since the loop
diagrams depend particularly on the values of $h H^+ H^-$
vertex, which is different for $V_A$ and $V_B$.
This study is complementary to the previous ones
\cite{BBS99,BS00}.

\section{Acknowledgment}
This work is supported by Funda\c{c}\~ao para a Ci\^encia e
Tecnologia under contract No. CERN/P/FIS/40131/00. S. O.
is supported by FCT under contract BM-20736/99.

\end{document}